\documentclass[review, authoryear]{elsarticle} \usepackage{comment}
\usepackage{lineno,hyperref}
\usepackage{hyperref}
\usepackage{comment}
\usepackage{graphicx}
\usepackage{makecell}
\usepackage{subcaption}
\usepackage[T1]{fontenc}
\usepackage[utf8]{inputenc}
\usepackage{float}

\newproof{pf}{Proof}
\newdefinition{rmk}{Definition}

\usepackage{amsmath}
\usepackage{amssymb}
\DeclareSymbolFontAlphabet{\mathcal}   {symbols}
\usepackage{graphicx}
\usepackage{algorithm}
\usepackage[noend]{algpseudocode}
\usepackage{caption}
\usepackage{subcaption}
\usepackage{multirow}
\usepackage{apalike}
\usepackage{hhline}
\usepackage[T1]{fontenc}
\usepackage[utf8]{inputenc}

\begin{document}

\begin{frontmatter}

\title{A Synthetic Network Generator for Covert Network Analytics}

\author[1,2]{Amr Elsisy}
\author[1,2]{Aamir Mandviwalla}
\author[1,2]{Boleslaw K. Szymanski\corref{mycorrespondingauthor}}
\cortext[mycorrespondingauthor]{Corresponding author}
\ead{szymab@rpi.edu}
\author[2,3]{Thomas Sharkey}

\address[1]{Department of Computer Science, Rensselaer Polytechnic Institute,  Troy, NY 12180, USA}
\address[2]{Network Science and Technology Center, Rensselaer Polytechnic Institute, Troy, NY 12180, USA}
\address[3]{Industrial and Systems Engineering, Rensselaer Polytechnic Institute, Troy, NY 12180, USA}

\begin{abstract}
We study social networks and focus on {\it covert} (also known as hidden) networks, such as terrorist or criminal networks. Their structures, memberships and activities are illegal.  Thus, data about covert networks is often incomplete and partially incorrect, making interpreting structures and activities of such networks challenging. For legal reasons, real data about active covert networks is inaccessible to researchers. To address these challenges, we introduce here a network generator for synthetic networks that are statistically similar to a real network but void of personal information about its members. The generator uses statistical data about a real or imagined covert organization network. It generates randomized instances of the Stochastic Block model of the network groups but preserves this network organizational structure. The direct use of such anonymized networks is for training on them the research and analytical tools for finding structure and dynamics of covert networks. Since these synthetic networks differ in their sets of edges and communities, they can be used as a new source for network analytics. First, they provide alternative interpretations of the data about the original network. The distribution of probabilities for these alternative interpretations enables new network analytics. The analysts can find community structures which are frequent, therefore stable under perturbations. They may also analyze how the stability changes with the strength of perturbation. For covert networks, the analysts can quantify statistically expected outcomes of interdiction. This kind of analytics applies to all complex network in which the data are incomplete or partially incorrect.

\end{abstract}


\end{frontmatter}

\section{Introduction}
The randomized generation of synthetic networks is often used for testing which properties of a given network depend on its structure and features \citep{barabasi2000scale,elsisy2019what}. Such a use has become popular since Erd{\'o}s and {R\'e}nyi introduced the random network model \citep{erdos1959random}. It generates an edge between each pair of nodes with a fixed probability $p$. The resulting networks are highly random as expected. Yet, despite interesting mathematical properties of this model, later research discovered that random networks rarely arise in nature or engineering practice. The more advanced models proposed later include the scale-free network model \citep{barabasi1999emergence} and its variants that represent the structures of many social and naturally arising networks. Another one is the Stochastic Block Model (SBM) \citep{holland1983stochastic} which extends the random network model by grouping nodes into blocks. The probability of an edge between a pair of nodes is now determined by the probability of connection between the blocks to which the nodes belong.  This model produces community structure resembling those arising in real networks but limits the differences between degrees of individual nodes located in the same block. This weakness is addressed by SBM variants, such as the degree planted SBM \citep{Karrer2011stochastic} or the degree-corrected planted partition model \citep{Newman2016equivalence}. The Lancichinetti-Fortunato-Radicchi (LFR) benchmark \citep{lancichinetti2008benchmark} generates synthetic networks with the desired heterogeneity of the node degrees and community size distribution. The generated networks are customized using such parameters as power law exponents for the node degrees, the community size, and the density of edges within the same community to mimic real networks. These networks are often used to test community detection methods. A model presented in \citep{xiao2020constructing} generates synthetic networks by rewiring edges in real-life networks. The more edges are rewired, the blurrier the communities become, allowing only the increasingly more stable communities to prevail. Another generative model creates a hierarchical multi-layer networks that are often used to represent the hierarchical management structures \citep{Ravasz2003hierarchical}. 

In general, we are interested in computer based social networks that have been supported by the growing number of computer platforms and companies. The activities in such networks generate massive amounts of data, to which, however, access is increasingly limited due to privacy concerns. We are particularly interested in covert (a.k.a. hidden) networks, such as terrorist and criminal networks. The common trait of such networks is that they try to hide their membership, structure, and illegal activities. The membership in such networks is often secretive. Their essential interactions are covert, so their non-essential interactions seem overly visible, making essential ones difficult to observe. Law enforcement in most of the countries is limited in the means of collected data by privacy laws protecting citizens from unjustified surveillance. As a result, the data about covert networks is often incomplete and partially incorrect. This creates a challenge in interpreting or discerning the structure and activities of such networks. An additional challenge arises from the inaccessibility to researchers of data about networks under investigations.  Only for a fraction of networks whose members were prosecuted in the court, such data becomes publicly available.

A network generator introduced here addresses these challenges. It is designed to create a set of synthetic networks, each structurally similar to a real network, but with anonymous nodes that are interconnected or clustered slightly differently than nodes in the original network. The direct use of such anonymized networks is to train research and analytical tools for finding structure and dynamics of covert networks. However, the synthetic networks provide alternative interpretations of the data about the original network. The distribution of probabilities for these alternative interpretations enables users to quantify statistically expected outcomes of operations on the covert networks.  Another use of the alternative interpretations is to decide if the data collected for the given covert network is sufficient for getting a reliable interpretation of this network's structure and dynamics.
A high frequency of alternative interpretations of the original network structure make them likely candidates for ground truth structure. To decide which interpretation is most likely the true structure of the original network may require collecting additional data. 

The synthetic networks created by our generator can also be useful for analyzing the social networks with partial information about their structures and bio-medical networks in which massive collection of experimental data about network dynamics makes the data partially incomplete and incorrect. 

\section{A model of organizations represented by covert networks}

Discovery and monitoring of covert networks often rely on getting access to information flows among nodes suspected to be involved in network activities. This flow may involve wiretapped telephone interactions, message exchanges, recorded conversations, or copies of written documents. We refer to an {\it organization} represented by such a network as {\it covert} and to the nodes as {\it members}. Using community detection, we discover what we call {\it groups} of members who interact among themselves more often than with other members. We prefer this term over {\it communities}, which usually denotes the nodes having casual relationships, or {\it clusters}, which refer to subsets of nodes with similar values for selected attributes. In small organizations, groups could be {\it independent} of the organizational structure of the network, but more often they have a {\it hierarchical} structure for management nodes. The number of hierarchy levels depends on the organization size. Our model represents each group by parameters such as the average density of edges inside the group, from this group to other groups, and the placements of group members in the organization hierarchy. The model uses parameters to individualize group leader connections to members of its group and, separately, to other nodes. Given an original  network, the generator individualizes it by randomly rewiring edges of its groups and its management hierarchy \citep{Ravasz2003hierarchical}. As a result, our model combines two network models: SBM for groups and hierarchical network for management structure.

Most of the previously proposed synthetic network generators rely only on one network generative model. For example, in \citep{shang2020growing} the authors propose a network generator that creates synthetic multi-layered networks. A network generator presented in \citep{guo2009random} implements small-world social networks with the desired high clustering coefficient. Another network generator described in \citep{benyahia2016dancer} creates dynamic networks with the prescribed community structure.

\section{Data}

To present our generator in action, we use the Caviar and Ciel datasets \citep{morselli2009inside}. 
The former defines a drug smuggling network, using data collected from 1994 to 1996 by an investigation of the West End Gang in Montreal, Canada. This gang was active in trafficking hashish and cocaine. During the investigation, police repeatedly confiscated shipments of drugs, but made no arrests until the investigation ended. The collected data mainly consists of the lists of phone calls made among gang members. 
Every two months, investigators were creating a snapshot of the network, in which each edge represents the calls made between two nodes during the corresponding two months and the edge weight is equal to the number of these calls. In total 11 snapshots were collected. The snapshots allow us to observe the network reactions to the shipment confiscations, and to the changes in positions of members in the network occurring between snapshots.

\begin{figure}[htpb]
\centering

\centering\includegraphics[width=10cm]{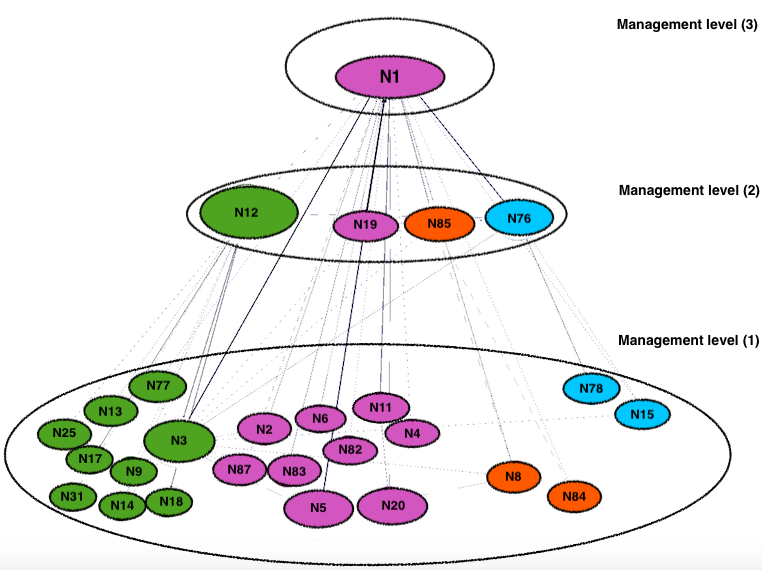}

\caption{Illustration of our model integration of the stochastic block model (SBM) with the hierarchical multi-layer network model in the Caviar network. Nodes marked with the same color belong to the same group. Nodes placed at the same hierarchy levels play the same roles in the management of this network.}
\label{fig:caviarHierarchy}
\end{figure}

In the case of the Caviar network, we know which nodes had some management roles from the publicly released court proceedings. For example, these proceedings identified node N1 (see Fig. \ref{fig:communitiesFigures}) as the leader of the hashish group, and node N12 as the leader of the cocaine group. Fig. \ref{fig:caviarHierarchy} shows more nodes in management roles in the Caviar groups. Still, the community detection algorithm alone did not assign many low degree nodes to any group \citep{bahulkar2018community}.
We also present the results obtained with the Ciel dataset \citep{morselli2009inside}, which defines an illicit drug transportation network that was engaged in trafficking in hashish from Jamaica in Montreal. The collection of data lasted from May 1996 to June 1997. The data defines a network with weighted edges representing the volumes of nodes' interactions obtained from the records of telephone calls and conversation surveillance. Only key managers of the network groups are identified. Nodes N1, N2, and N10 all took part in leading the network. There was no listing of the ground truth groups. For both networks, we use the Louvain community detection algorithm \citep{blondel2008fast} to find groups and the relative betweenness centrality to find nodes involved in managerial roles. Since criminal networks often have sparse connectivity because of attempts to hide network activities, we execute a version of Louvain algorithm for graphs with undirected edges. For this execution, we temporally  transform directed edges of the generated networks into undirected ones, summing their weights for pairs of nodes which have two opposing edges connecting them. 
\begin{table}[htpb]
\centering
\begin{tabular}{|c|c|c|c|c|c|c|c|}
\hline
\thead{Network} & \multicolumn{4}{c|}{Caviar}&\multicolumn{3}{c|}{Ciel} \\
\hline 
\thead{\small Node identifier} &\small N1 &\small N3 &\small N12 &\small N76 &\small N1 &\small N2 &\small N10 \\ 
\hline
\thead{\small RBC score} &\small 0.430 &\small 0.180 &\small 0.303 &\small 0.078 &\small 0.591 &\small 0.641 &\small 0.015\\ \hline
\thead{\small Rank of the score} &\small 1 &\small 3 &\small 2 &\small 4 &\small 2 &\small 1 &\small 7\\

\hline

\end{tabular}
\caption{\label{tab:Betweenness} The relative betweenness centrality of the management nodes of the Caviar and Ciel networks reveals their different priorities in managing criminal organization. In the Caviar network managers are strongly connected to the nodes in their groups, which improves efficiency of management. In contrast, In Ciel network management nodes connect direct only to subordinates. This may benefit security of node N10.}
\end{table}
Covert networks may prioritize either efficiency or security, but not both. The betweenness centrality of a node measures a fraction of the shortest paths of information flow between all pairs of nodes passing through this node \citep{Unnithan2014Betweennes}.  
A normalized version of this metric, called relative betweenness centrality, limits its range to [0,1]. We use it for ease of comparison of the results. We conclude that the structures of the Caviar and Ciel networks are fundamentally different. The Caviar network prioritized efficiency. This is indicated by the highest relative betweenness centrality scores of the management nodes, which are N1, boss, and N3, N12, N76, managers, among all nodes in the Caviar network (cf. Table \ref{tab:Betweenness}). The implied easy access to these nodes from others supports high efficiency of information flow in the network, but low security for the leader nodes. In contrast, in the Ciel network, the boss limits the connections to direct subordinates. Consequently, node N10 has a very low relative betweenness centrality (RBC) when compared to the manager nodes, N1 and N2, (see Table \ref{tab:Betweenness}) even though some management role was plausible for him \citep{morselli2009inside}. 

We chose the Caviar and Ciel datasets for our study because they are well studied and analyzed using public data released during court proceedings \citep{bahulkar2018community, berlusconi2013all, skillicorn2013spectral}. We show that synthetic networks generated by our generator for both networks are similar to the original ones. This is important because only pieces of the ground truth are available for covert networks under investigation. 

\section{Implementation of the Random Anonymized Network Generator}

\subsection{Overview}
The {\it Random Anonymized Network Generator} (RANG) organizes its network generation process into the following three steps. The owners of the original network data execute the first step. They need to replace node identity and its personal data by a single unique numerical ID of this node. They also need to assign to each node its place in the management hierarchy, a group to which this node belongs and two lists, one of the subordinates and the other of superiors. The second step is executed by RANG software. It summarizes the group structure of the network as the list of probabilities of an edge between any pair of nodes. These probabilities depend on groups to which the nodes belong and the roles these nodes play in the network. The obtained data are shared with the outside users or used internally by the owners. The third step generates a set of synthetic random networks using the edge probabilities defined in the second step and analyzes this set for a group structure stability and nodes' management roles consistency.  

\subsection{Generating groups in covert networks}

Social networks often have directed weighted edges to represent intensity of interactions measured in frequency of calls, messages, or meetings. Within SBM, we use one of the two models to assign weights to edges. 

The first process for assigning edges is called the Weighted Random Graph (WRG) generator \citep{Garlaschelli2009weighted}. Let $W$ denote the sum of weight of all edges and $E$ the maximum number of edges we can generate between subsets of nodes. In WRG, the edges' weights are generated by running Bernoulli trials with probability $p=\frac{W}{W+E}$. This run stops at the first failed trial. The number of successful trials before this failure defines the weight of the generated edge. This process gives rise to the geometric distribution of edge weights. 

We introduce here an alternative approach, named {\it Bernoulli Weighted Random Network} (BWRN) model. It has two parameters. One is a vector of the weights $w$'s for edges in the original graph, and the other is the probability $p_B$ that controls the variance of the generated weight distribution. The process of generating edges starts with the heaviest edges and progresses down to edges with the smallest weight. 
Given the currently processed weight $w$, an associated weight is computed as $w_B=\lfloor w/p_B \rfloor$. For each edge with $w$ weight in the original graph, we select a weight in the range $[0,\lceil w/p_B \rceil]$ as follows. First, we select randomly a pair of not yet connected nodes and run $w_B$ Bernoulli trials with probability $p_B$. If $p_Bw_B<w$, we do one more Bernoulli trial with probability $p_a=w-p_Bw_B$. The weight from such a run is equal to the number of successes in those trials. The edge is not created when this run returns weight 0.

This design yields a distribution of weights with probability of choosing weight $k$, where $0\leq k\leq \lceil w/p_B\rceil$ is defined as:
\begin{equation} 
p_w(k,p_B)=\left\{ 
                \begin{array}{ll}
                    \frac{(w_B)!}{(w_B-k)!k!}p_B^{k}(1-p_B)^{w_B-k}& \mbox{if $k\leq w_B$}\\
                    w-p_Bw_B & \mbox{for $k=w_B+1$ if $p_Bw_B<w$}
                    \end{array}
           \right.
\label{eq:bwrn}
\end{equation}

As a result, the average sum of weights of all edges created by this process is the same as in the original network. Indeed, the expected weight from Equation \ref{eq:bwrn} is $p_Bw_B+w-p_Bw_B=w$. 

The parameter $p_B$ defines probability that the edge of weight $w$ will not be generated, which is $(1-p_B)^{w_B}$ if $w=p_Bw_B$ and
$(1-p_B)^{w_B}(1-w+p_Bw)$ otherwise, so it quickly decreases with increase of $w$ and $p_B$. With $p_B>0.9$ even edges with weight 1 have low chance of below $1\%$ to be lost. An interesting trade-off arises for slightly lower values of $p_B$. For example with $p_B=0.875$, which we used for computational experiments here, about $10\%$ of such edges will not appear in the generated network but a similar fraction of edges will increase their weight to 2, strengthening cohesiveness of some communities. 

The variance of the distribution of the weights generated for an edge with weight $w$ in the original data is $w(1-p_B)+p_a(p_B-p_a)\approx w(1-p_B) $~\citep{feller1968}, so it grows with increase in $w$ but decays with increase of $p_B$\footnote{Indeed $\sum_{k=0}^{w_B} k^2p_k(1-p_a)+k^2*p_kp_a+(2k+1)p_kp_a = w(1-p_B)+w^2+p_a(p_B-p_a)$. Hence $Var(p_B,w)=w(1-p_B)+p_a(p_B-p_a)$}. Thus, selecting large $p_B$ will make generated synthetic networks more similar to the original one, while decreasing it would have opposite effects. Hence, different kinds of analyses could be conducted with different choices of $p_B$. 

By taking into account weights of edges, our model allows users to define different edge densities in the group for edges at the same level of hierarchy (e.g., among peers) than for edges across the hierarchy (e.g., between the group leader and a subordinate). This enables users to account for typically higher information flow intensities between managers and subordinates than among peers.

\subsection{Detecting groups in the generated networks}

To test the accuracy of our synthetic network generator, we ran the Louvain community detection algorithm \cite{blondel2008fast} on all the generated networks and compared the generated groups to the groups in the original network. Yet, there are many high quality community detection algorithms, such as SpeakEasy~\cite{speakeasy}, which we triesd and it yielded similar results, CPM \cite{Traag2011narrow}, modularity maximization \cite{chen2014community}, fast \cite{Clauset2004Finding}, or adaptive modularity \cite{Lu2018Adaptive} that can be used for this purpose on social and covert networks, so users of RANG can use any of them for this purpose.

We observe that in the Caviar and Ciel networks, these two groups often do not match perfectly. There are several reasons for such differences. First, datasets for covert networks are incomplete and may have many undetected edges. Second, important nodes are often so highly interconnected, that they may belong to several different groups. For example, in Caviar, nodes N1 and N3 belong to the same group in the originla graph, but N3 has so many connections to other groups that the generator occasionally assigns it to them. For similar reasons such miss-assignments may happen to the managers. At the same time, such a miss-assignment may signal diverse roles that such a node plays in the network. Hence, comparative analysis of groups discovered in the synthetically generated networks may shed a new light on the operations of the original covert network.

We also detect the network hierarchy levels to reveal the structural properties of the generated networks. We use the relative betweenness centrality, which is easy to compute, as a hierarchy level measure because in networks there is a strong correlation between the hierarchy measures and betweenness centrality scores \citep{rajeh2020interplay}.
Comparing the nodes with high relative betweenness centrality in the generated network and such nodes in the original network enables us to measure how well the generated networks preserve leadership hierarchy. A Combined Score (CS) measures overall similarity of generated networks to the original one. CS is a product of the group and hierarchy similarities.

\subsection{Steps for generating synthetic networks}
The owners of the data about the original (real or synthetic) network need to provide as input three lists about this network. They need first to anonymize input data by removing any personal information about the nodes and instead provide just lists of nodes, each represented by a pair with an abstract node identifier and the level of management hierarchy to which this node belongs. The second is the list of all edges, each providing identifiers of its starting and ending nodes and its weight. The third is the list of groups, each group represented by the list of identifiers of its  members based on the original network and, unless a group is independent, the identifier of a group leader. Alternatively, the owners may prefer to generate a synthetic network from this input since the output of the generator is in the required format and the generated network is different from the original.

The processing of the input data starts with computation of probabilities of existence and weights of edges for the generated network. We use two models to assign randomized weights to the weighted edges. Both methods classify the edges into several classes and generate edges and weights for each class separately. The first class includes internal edges of a group $g_i$ of size $|g_i|$ at the same level of management hierarchy, so including members but not superiors of the neighbors. In such a case, there are $E^i=|g_i|(|g_i|-1)$ directed edges. Let $W^i$ denote the sum of their weights. The second class includes edges across the members of two different groups $g_i, g_j$. There are $E^{i,j}=|g_i||g_j|$ such edges from $g_i$ to $g_j$ and $E^{j,i}$ in the opposite direction. Let sums of their weights be denoted as $W^{i,j}, W^{j,i}$. The next class includes edges from a superior, $s(i)$ to the members of its group $g_i$ and from the group members to this superior. There are $E^{s(i),i}=E^{i,s(i)}=|g_i|$ of such edges in each direction. Their sums of weights are denoted as $W^{s(i),i},W^{i,s(i)}$. Let $|\neg i|$ denote the number of all nodes not in a group $g_i$ at the level of management of members of this group. We define the class of edges from the superior of group $i$ to $\neg i$ nodes as $E^{s(i),\neg i}=|\neg i|$ in each direction, with the sum of weights denoted $W^{s(i),\neg i}$. 

One solution uses the Weighted Random Graph (WRG) approach \citep{Garlaschelli2009weighted}.  
The second solution uses the Bernoulli Weighted Random Network (BWRN).
Section {\it Generative models networks} discusses both solutions.

Next, the management roles are assigned to the nodes.
Our model allows for an arbitrary number of management hierarchy levels. Yet, given that both Caviar and Ciel networks are composed of small groups, we set here the limit for their hierarchy levels at three. The number of nodes assigned to each hierarchy level higher than one depends on the total number of groups at the immediately lower level. The third level of management consists of the highest authority node in the local network. We refer to such a node as a {\it boss}. We refer to the nodes at the second level as {\it managers}. The first level of hierarchy is composed of the remaining nodes organized into groups. We call such nodes {\it members}. As explained in the next section, our network generator attempts to find the group managers even without any information from the network investigators. Managers serve as intermediaries between the boss and the members. The review of the literature reveals that the majority of small companies have a ratio of four employees per manager \citep{davison2003management}. This is a bit smaller than in case of covert networks, such as Caviar, where this ratio is close to six. The plausible reasons for this difference include self-motivation of the members for doing their tasks, as well as keeping a fraction of the organization members interacting with the boss small for safety reasons. 

Not all groups are supervised through hierarchical management. We refer to such a group as {\it independent}. A group of small size and a low fraction of reciprocal connections is likely to be independent. An example is the money-laundering group in the Caviar network that might be regarded as independent. Its members have only outgoing edges targeting the outside nodes. Thus, members do not have incoming edges from any other node in the network.

\subsection{Baseline network generator}\label{subs:Baseline}
To get a better baseline for network accuracy than the Erd{\"o}s-R{\'e}nyi random network, we built a straightforward extension of SBM by supporting weighted directed edges with integer weight. It takes as input lists of network nodes, edges with integer weights, and groups in the original network.  The first step is to count, for each group, the sum of weights of all edges inside this group, and then edges of this group leading to and from each other group. These sums become probabilities by dividing them by the sum of weights of all the network edges. Using the built-in numpy~\citep{oliphant2006guide} random choice method, the baseline generator picks a pair of groups, including those in which the source and target groups are the same. Then, a node from the source group and a node from the target group are selected repeatedly until the selected nodes are different, thus avoiding self-loops.  Using the created set of probabilities, we select the probability for the selected group and execute a Bernoulli trial with this probability. On success of the trial, the weight of connection between these two nodes is increased by one.  This entire process is repeated until the total weight of all edges becomes the same as in the input network.  The last step is to run Louvain community detection on the resulting network and compare the output with the original network communities.

\section{Results}
We test the network generator by generating 100 synthetic networks and taking the average scores to avoid any statistically insignificant outliers. We use the Louvain community detection algorithm to find the group structure of each generated network. Then, using the Normalized Mutual Information (NMI) method \citep{danon2005comparing}, we compare groups detected in the generated networks to groups in the corresponding original network.

\subsection{Results with Caviar and Ciel datasets}

\begin{table}[htpb]
\centering
\begin{tabular}{|c||c|c|c||c|c|c||}
\hline
\thead{Original Network}&\multicolumn{3}{|c||}{\thead{Caviar}}&\multicolumn{3}{|c||}{\thead{Ciel}}\\
\hline
\thead{Generator}&
\multicolumn{2}{c|}{\thead{RANG}}&\thead{Weighted}&\multicolumn{2}{c|}{\thead{RANG}}&\thead{Weighted}\\
&\thead{BWRN}&\thead{WRG}&\thead{SBM}&\thead{BWRN}&\thead{WRG}&\thead{SBM}\\
\hline
\thead{mean}   & 0.815     & 0.356  &\bf 0.850 & \bf 0.800 & 0.513 & 0.761 \\
\hline
\thead{median} & \bf 0.839 & 0.365 & 0.739 &\bf 0.883  &  0.567 & 0.751\\
\hline
\thead{min} & 0.415 & 0.088 & 0.358  & 0.734 & 0.288 & 0.551 \\
\hline
\thead{max} & 0.953 & 0.565 & 0.883  & 1.000 & 0.692  & 1.000\\
\hline

\end{tabular}
\caption{\label{tab:nmiCaviarCiel} 
Results show NMI scores from comparing the groups in networks generated from the Caviar (columns 2-4) and Ciel (columns 5-7) original networks to groups in those original networks. Three generators are compared. Columns 2, 5, show performance of RANG with our Bernoulli Weighted Random Generator. In Columns 3, 6 results for RANG using Weighted Random Graph method are displayed. Columns 4-7 contain results for the weighted SBM baseline method. Our BWNR method outperformed the other two in terms of median values for both Caviar and Ciel networks and mean value for Ciel network. The weighted SBM method was close behind with the best score for mean value for Caviar. WRG method was far behind with scores about half of the leading scores. } 
\end{table}

To evaluate how much the RANG generated synthetic networks are similar to the original network used for input to RANG, we measure two aspects of similarity. The first is group structure. For this purpose, we used Louvain community detection to find group structure of all compared networks. The second aspect is management nodes. We find all management nodes by their high relative betweenness centrality in all compared networks. We use as threshold $90\%$ of the relative betweenness centrality metric of the node whose rank among the highest values of this metric is equal to the number of management nodes in the original network. For both datasets, we used 
weighted edge generators to create three sets of synthetic networks. 
For the first set we used our Bernoulli Weighted Random Network method, for the second set, the Weighted Random Graph method, while for the last set, the weighted SBM described in Section \ref{subs:Baseline}.  

Table ~\ref{tab:nmiCaviarCiel} contains results of measurements of the first aspect of similarity that synthetic networks generated by BWRN methods are by factor of two more similar to original group structures than groups in networks generated with WRG method. The BWRN synthetic model and the weighted SBM baseline performed close to each other. 

\begin{table}[htbp]
\centering
\begin{tabular}{|c|c|c|c|}
\hline
\thead{Metric}&\thead{RANG \& BWRN}&\thead{RANG \& WRG}&\thead{Weighted SBM} \\
\hline
\thead{Group NMI median}          &\bf 0.839 & 0.365 & 0.739\\
\hline
\thead{Jaccard Leadership}        &\bf 0.681 & 0.402 & 0.308\\
\hline
\thead{Combined Score}            &\bf 0.571 & 0.146 & 0.281\\
\hline

\end{tabular}
\caption{\label{tab:LeadershipDetection}
Similarity of the original Caviar network to networks generated by our Random Anonymized Network Generator first with our Bernoulli Weighted Random Network method, then with Weighted Random Graph method. The third set of networks was generated by Weighted SBM baseline. The first row of the results shows group structure similarity from Table \ref{tab:nmiCaviarCiel}, the second the leadership similarity, and the last row shows the combined Score that is the product of the first two scores. In all three rows, the best score is shown in bold font.}  
\end{table}
This means that the management structure is a differentiating factor in comparing these two generators. It is evaluated in Table~\ref {tab:LeadershipDetection} that demonstrates the importance of leadership detection for keeping the synthetic distinct but close to the original network.
The results obtained using Jaccard metric \citep{zaki2020} 
show that the RANG generated networks using our Bernoulli Weighted Random Network method preserve the group structure and leadership hierarchy twice as well as the baseline generated networks and nearly four times better than RANG using Weighted Random Graph method. Moreover, the unweighted SBM generator performs much worse than the baseline with the weighted SBM. 
\begin{figure}[htpb]
\centering
\vspace{-0.2cm}
\begin{subfigure}{6cm}
\includegraphics[width=4.5cm, height=5cm]{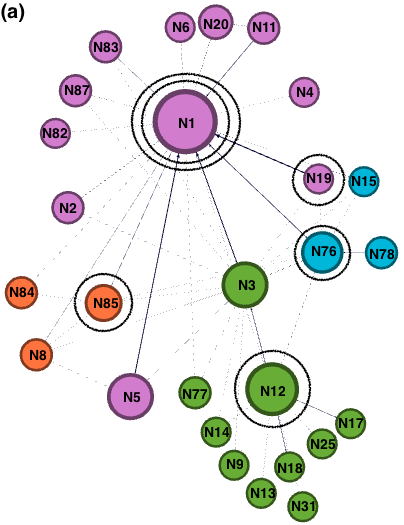}
\end{subfigure}
\begin{subfigure}{6cm}
\includegraphics[width=4.5cm, height=5cm]{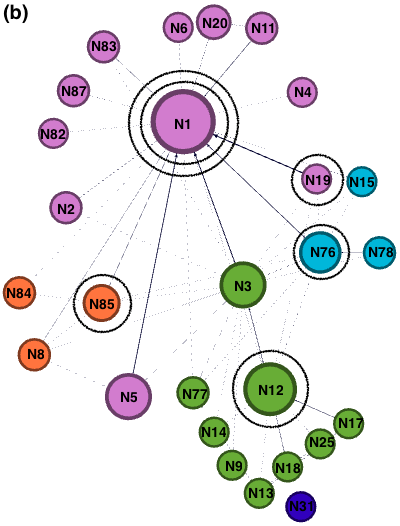}
\end{subfigure}\vspace{15pt}

\begin{subfigure}{6cm}
\includegraphics[width=4.5cm, height=5cm]{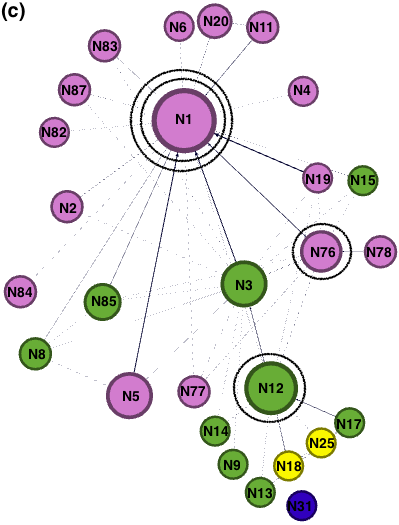}
\end{subfigure}%
\begin{subfigure}{6cm}
\includegraphics[width=4.5cm,height=5cm]{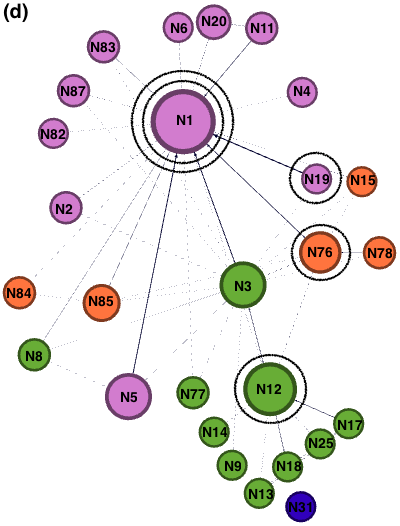}
\end{subfigure}

\caption{The presented Caviar networks depict a) the groups  in the original network, and groups detected in the synthetic networks with similarity that is (b) highest, (c) lowest, and (d) average. 
}
\label{fig:communitiesFigures}
\end{figure}
n summary, our RANG approach better differentiates between leadership and membership roles for nodes within groups while performing at least slightly better than others on community detection. These results verify the strength of RANG in preserving groups and hierarchy within the generated networks.

\subsection{Stability of group structures of covert networks}
Generating statistically similar networks to a real network enables us to analyze how stable is the structure of the original network. These networks represent sets of small perturbations of the original network interconnections. If many generated synthetic networks are structurally similar to the original network, the latter network is stable and resistant to such perturbations. On the other hand, if the structure of the original network is not stable, only a few synthetic networks will be structurally similar to the real network. 

\begin{figure}[htpb]
\centering

\centering\includegraphics[width=10cm]{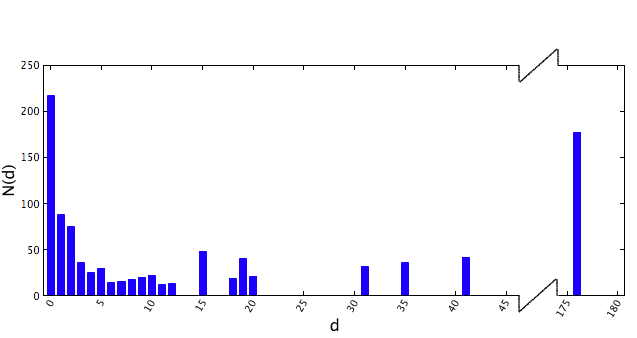}

\caption{The histogram of 
meta-graph degree distribution for exact matching of groups in the generated networks. The x-axis defines the node degree $d$, while the y-axis shows $n (d)$, the number of nodes with $d$ degree. The group structure of the original network is not stable because only $1\%$ of synthetic networks share this structure.}
\label{fig:generatedNetworksStability}
\end{figure}

We apply this analysis to the Caviar network, by generating $G =1000$ random synthetic networks, and comparing their structures to each other, and to the structure of the Caviar original network. We start this analysis by creating a meta-graph, nodes of which are the generated networks, so the size of the meta-graph is $G=1000$ nodes. For each node, we draw an undirected edge between this node and any other node in the meta-graph with a matching group structure. We create two versions of the meta-graph. In one, edges represent an exact matching. In another, the edges show flexible matching, which allows up to one node difference in each group for drawing an edge. Figure \ref{fig:generatedNetworksStability} shows the resulting distribution of the node degrees in meta-graph with exact matching.
We find that the original network's structure repeats ten times for exact matching among 1000 generated synthetic networks. Thus, this structure is not very stable. It is sensitive to small perturbations in node connectivity in the generated networks. In contrast, we also find that the most frequent group structure occurs 177 times with exact matching, and 310 times for flexible matching. It also happens to be the most similar to the original detected group structure. The synthetic network with this structure is shown in Sub-figure \ref{fig:communitiesFigures}(b). 
Moreover, each of the top ten most frequently occurring structures repeats at least 20 times for exact matching, and at least 206 times for flexible matching. The remaining generated networks have either unique structures, or structures that were similar to only a few generated networks.  

From a practical perspective, the identification of stable groups in a criminal organization is important. It will help analysts to concentrate on group structures that arise frequently and thus represent plausible interpretations of data collected about the network. Simulating interdiction on alternative structures reveals a range of outcomes.  Using the meta-graph of generated networks, analysts can compute probabilities of these outcomes providing the distribution of the interdiction outcomes for the given original network.

\section{Conclusions}

We introduce and make available to researchers and analysts a network generator that produces synthetic networks statistically similar to the given real or synthetic network. The first goal of this work is to enable sharing data between analysts who own sensitive data about current investigations and researchers, who need realistic networks to develop efficient tools for network analytics. Only owners of the data can enable sharing of sensitive data. They can properly separate abstract structural information about network from the sensitive personal and operational information. The former is first encapsulated by owners into three lists of nodes, groups and management roles and then further transformed into numeric, possible shuffled node ids, nodes' clustering into groups and statistics about edges and their weights across those clusters. Hence, the description of this original networks passed to researchers is succinct and void of personal data. The output of the generated networks follows the same format as is required for input. The model uses the introduced here Bernoulli Weighted Random Network generator that creates the edges whose average sum of weights is the same as it is in the original network. It also uses the Stochastic Block Model to create alternative group structures to those existing in the original network. To preserve the organizational aspects of the original network, we use a hierarchical network model.

In general, our generator is broadly applicable to social networks, both with formal and informal organizational structures. It aims at aiding analyses and investigations of covert networks with partial information about their nodes, edges, and internal structures.
The generator enables researchers to develop algorithms and systems on the generated synthetic networks for use in real applications.

Even the analyst have only partially knowledge of the covert networks during investigation. Therefore, the important second use of the generator is as a tool for analysts to use it for network analytics on the original network. Studying the generated synthetic networks statistically similar to a given covert network is useful in two ways. Differences between those networks suggest the potential alternative interpretations of data or the need for collecting more data. The similarities between those networks will enable the analysts to find stable parts of the original network structure. The more synthetic networks we have, the more we can analyze and understand the operations, leadership, and groups present in the original network. More generally,
the distribution of probabilities for the alternative interpretations enables new network analytics. The analysts can find community structures which are frequent, therefore stable under perturbations. They may also analyze how the stability changes with the strength of perturbation. For covert networks, the analysts can quantify statistically expected outcomes of interdiction. This kind of analytics applies to all complex network in which the data are incomplete or partially incorrect.

We thoroughly tested the generator on two real covert networks, Caviar and Ciel. To measure the similarity between the generated networks and the original one, we use the well-known Louvain community detection algorithm. Applying the NMI and the Jaccard metrics, we measured the results, which demonstrate the high similarities among generated networks and the original one. 

We conclude that accounting for groups and a management hierarchy of a network is essential for generating synthetic networks that are statistically similar to the original network.

In future work, we plan to add the fourth step of network generation going beyond the original graph to enable its {\it hierarchical network expansion}. One way to accomplish it is to replicate any part of the original network multiple times at any level of hierarchy and provide additional levels of management hierarchy if needed. Such an extension will allow the researchers and analysts to study evolution of covert networks and potential interdiction outcomes in large and complex criminal networks.

\bibliography{main}

\section{Author contributions statement}
\textbf{Amr Elsisy}: Conceptualization, Methodology, Software, Investigation, Writing - Original Draft, Writing - Review \& Editing, Visualization. \textbf{Aamir Mandviwalla}: Methodology, Software, Investigation, Writing - Original Draft, Writing - Review \& Editing. \textbf{Boleslaw K. Szymanski}: Conceptualization, Methodology, Writing - Original Draft, Writing - Review \& Editing, Supervision, Project Administration. \textbf{Thomas Sharkey}: Writing - Review \& Editing.

\section{Acknowledgments}
This work was partially supported by the U.S. Department of Homeland Security under Grant Award Number 2017-ST061-CINA01, the Office of Naval Research (ONR) Grant N00014-15-1-2640, the Defense Advanced Research Projects Agency (DARPA) and the Army Research Office (ARO) under Contract W911NF-17-C-0099, and the Army Research Office (ARO) Grant  W911NF-16-1-0524. 
The views and conclusions contained in this document are those of the authors and should not be interpreted as necessarily representing the official policies, either expressed or implied, of the U.S. Department of Homeland Security or U.S. Department of Defense.

\section{Competing Interests}
The authors declare no competing interests.

\end{document}